\documentclass[twocolumn,pra,superscriptaddress]{revtex4}
\usepackage{amsmath,amssymb,}
\usepackage{graphicx}
\usepackage{dcolumn}
\usepackage{bm}

\usepackage[colorlinks,citecolor=red]{hyperref}

\usepackage{rotating}
\usepackage{floatrow}

\allowdisplaybreaks[2]

\begin{document}

\title{Double exceptional links in a three-dimensional dissipative cold atomic gas}

\author{Peng He}
\author{Jia-Hao Fu}
\affiliation{National Laboratory of Solid State Microstructures
and School of Physics, Nanjing University, Nanjing 210093, China}

\author{Dan-Wei Zhang}
\email{danweizhang@m.scnu.edu.cn}
\affiliation{Guangdong Provincial Key Laboratory of Quantum Engineering and Quantum Materials, GPETR Center for Quantum Precision Measurement and SPTE, South China Normal University, Guangzhou 510006, China}

\affiliation{Frontier Research Institute for Physics,
South China Normal University, Guangzhou 510006, China}

\author{Shi-Liang Zhu}
	\email{slzhu@nju.edu.cn}
\affiliation{National Laboratory of Solid
State Microstructures and School of Physics, Nanjing University,
Nanjing 210093, China}

\affiliation{Guangdong Provincial Key Laboratory of Quantum Engineering and Quantum Materials, GPETR Center for Quantum Precision Measurement and SPTE, South China Normal University, Guangzhou 510006, China}

\affiliation{Frontier Research Institute for Physics,
South China Normal University, Guangzhou 510006, China}

\date{\today}

\begin{abstract}
We explore the topological properties of non-Hermitian nodal-link semimetals with dissipative cold atoms in a three-dimensional optical lattice. We construct a two-band continuum model in three dimensions with a spin-dependent gain and loss, where the exceptional points in the energy spectrum can comprise a double Hopf link. 
The topology of the bulk band is characterized by a winding number defined for a one-dimensional loop in the momentum space and a topological transition of the nodal structures emerges as the change of the non-Hermiticity strength. A non-Bloch theory is built to describe the corresponding lattice model which has anomalous bulk-boundary correspondence. Furthermore, we propose that the model can be realized by using ultracold fermionic atoms in an optical lattice
and the exceptional nodal links as well as the topological properties can be detected by measuring the atomic spin textures.
\end{abstract}

\maketitle

\section{Introduction}
Topological metals and semimetals with gapless bulk nodes protected by certain symmetries have attracted intense interest both theoretically and experimentally in the past decades \cite{Chiu2016,DWZhang2018}. Point and line nodes such as Dirac, Weyl and Maxwell points \cite{SLZhu2007,Young2012,ZWang2012,BLv2015,Young2015,Armitage2018,YQZhu2017,XTan2018,XWan2011,XSLi2019,MXDeng2019}, nodal lines \cite{Burkov2011,Fang2015,FangT2016,Yu2015,Yan2016,DWZhang2016}, and nodal links and knots \cite{Bzdusek2016,Sun2017,YanNLS2017,Chen2017,Chang2017,Ezawa2017,Bi2017} in gapless phases are topological defects in momentum space carrying topological charges. Such degeneracies occur close to the Fermi level, such that the stability of the Fermi surface and the low-energy excitations of the semimetals are topologically protected. The stability of the Fermi surface originates from the topological properties of Green's function \cite{Volovik2003}. The inclusion of a self-energy in Green's function naturally necessitates a non-Hermitian Hamiltonian \cite{Zirnstein2019,Borgnia2020,Hirsbrunner2019}, emerging as an effective description of a nonconserved system such as solids with finite quasiparticle lifetimes \cite{Kozii2017,Yoshida2018,Yoshida2019,Kimura2019,Michishita2020}, disordered Dirac fermions \cite{Zyuzin2018,Papaj2019}, and artificial lattices with gain and loss or nonreciprocity \cite{Klaiman2008,Cerjan2019,Zhou2018,JLi2019,WGuo2020}.

In recent years, there has been considerable attentions on the topologies of non-Hermitian systems \cite{BergholtzET2019,Ghatak2019,HJiang2019,DWZhang2020,XWLuo2020,HWu2020,HLiu,LZTang,KunstBio2018,ZWang2018,KawabataS2019}. A framework resting on the universal Green's function has been formulated to describe anomalous topological characteristics of the non-Hermitian Hamiltonians \cite{Zirnstein2019,Borgnia2020}. The eigenenergies of a non-Hermitian Hamiltonian are generally complex and the gaps are defined on the complex manifold. This enables two types of gaps: line gap and point gap \cite{KawabataS2019,KawabataC2019,OkumaTopo2019}. Some unique topological phenomena such as the non-Hermitian skin effect \cite{ZWang2018,FSong2019,LiGong2019,CHLee2019,LeeHybrid2019,Hofmann2019,OkumaTopo2019,CHLeeUn2019}, the anomalous bulk-boundary correspondence \cite{ZWang2018,YaoNHChern2018,KunstBio2018,Murakami2019,CFang2019,LLiC2020,XRWang2020}, and the exceptional points can be attributed to the point-gap topology. Furthermore, as pinpointed in previous works \cite{ZhangHJ2019,Hu2019,Bergholtz2019}, the nodal band structures for non-Hermitian Hamiltonian depend only on the topology and do not require any protecting symmetries. In contrast, the topological semimetals in the absence of non-Hermiticities can only be protected by symmetries due to the so-called band repulsion \cite{FangT2016}. Thus, the nodal points for non-Hermitian Hamiltonians can host new topology without Hermitian counterparts, such as Weyl exceptional rings with a Chern number \cite{XuDuan2017}, nodal rings \cite{ZhangHJ2019,Kunst2019,Yamamoto2019} and exceptional knots and Hopf links \cite{Bergholtz2018,Hu2019,ZYang2020,Bergholtz2019}.

On the other hand, the experimental demonstration of non-Hermitian classical and quantum systems has witnessed great progress. Particularly promising approaches using different platforms such as acoustic \cite{ZhuZhang2014,Popa2014}, optical \cite{Zhou2018,Cerjan2019,Xiao2019,KWang2019} or atomic \cite{SDiehl2011,JLi2019,Takasu2020} systems have been successfully performed. The present experimental achievements have extended the field of searching for exotic topological phases with synthetic quantum matter to the unconventional non-Hermitian situation by engineering the dissipation. Specifically, several proposals based on the optical lattices has been proposed \cite{XuDuan2017,Yamamoto2019,Ashida2017,Gong2018,LiGong2019}.

In this work, we explore the topological properties of a non-Hermitian three-dimensional (3D) nodal-link semimetal. The non-Hermiticity considered here is introduced only by a spin-dependent gain-and-loss term that can be easily realized \cite{SDiehl2011,JLi2019}, while exceptional links or knots in non-Hermitian metals or semimetals proposed in previous works \cite{Bergholtz2018,Hu2019,Bergholtz2019} usually arise from delicately designed non-Hermitian Hamiltonians. We find a double Hopf link composed of exceptional degeneracies for some regimes of parameters, and a topological transition of the nodal structure emerges as the change of the non-Hermiticity. We use a winding number defined on an $\mathcal{S}^1$ sphere that encloses the nodal rings to characterize the topological properties of our system.
Moreover, a non-Bloch band theory is built to restore the bulk-boundary correspondence in this 3D non-Hermitian system, and a critical phenomenon is observed. Specifically, we find that the topological zero-energy edge modes are gapped out for small systems, distinguishing then from conventional Hermitian systems in which the finite size effects play a diminishing role. Furthermore, we propose a scheme to realize our model by using the two-photon Raman assisted tunneling of ultracold fermionic atoms in an optical lattice \cite{Jaksch2003,HMiyake2013,Aidelsburger2013,DWZhang2018}. Motivated by a recent work that realized a nodal-line semimetal using spin-orbit coupling of cold atoms \cite{Song2019}, we propose an approach to detect the link structure of the exceptional degeneracies by the integrated spin textures.

This paper is organized as follows. Sec. \ref{sec2} introduces a non-Hermitian continuum model exhibiting double exceptional links in the energy spectrum and the topological properties of this system are discussed. In Sec. \ref{sec3}, the corresponding lattice modal is constructed and the anomalous bulk-boundary correspondence of this system is addressed. We further propose the realization and detection of topological exceptional links in optical lattices in Sec. \ref{sec4}. Finally, a short conclusion is given in Sec. \ref{sec5}.

\section{A continuum model with double exceptional links}\label{sec2}

\begin{figure}[htbp]
	\centering
	\includegraphics[width=\textwidth]{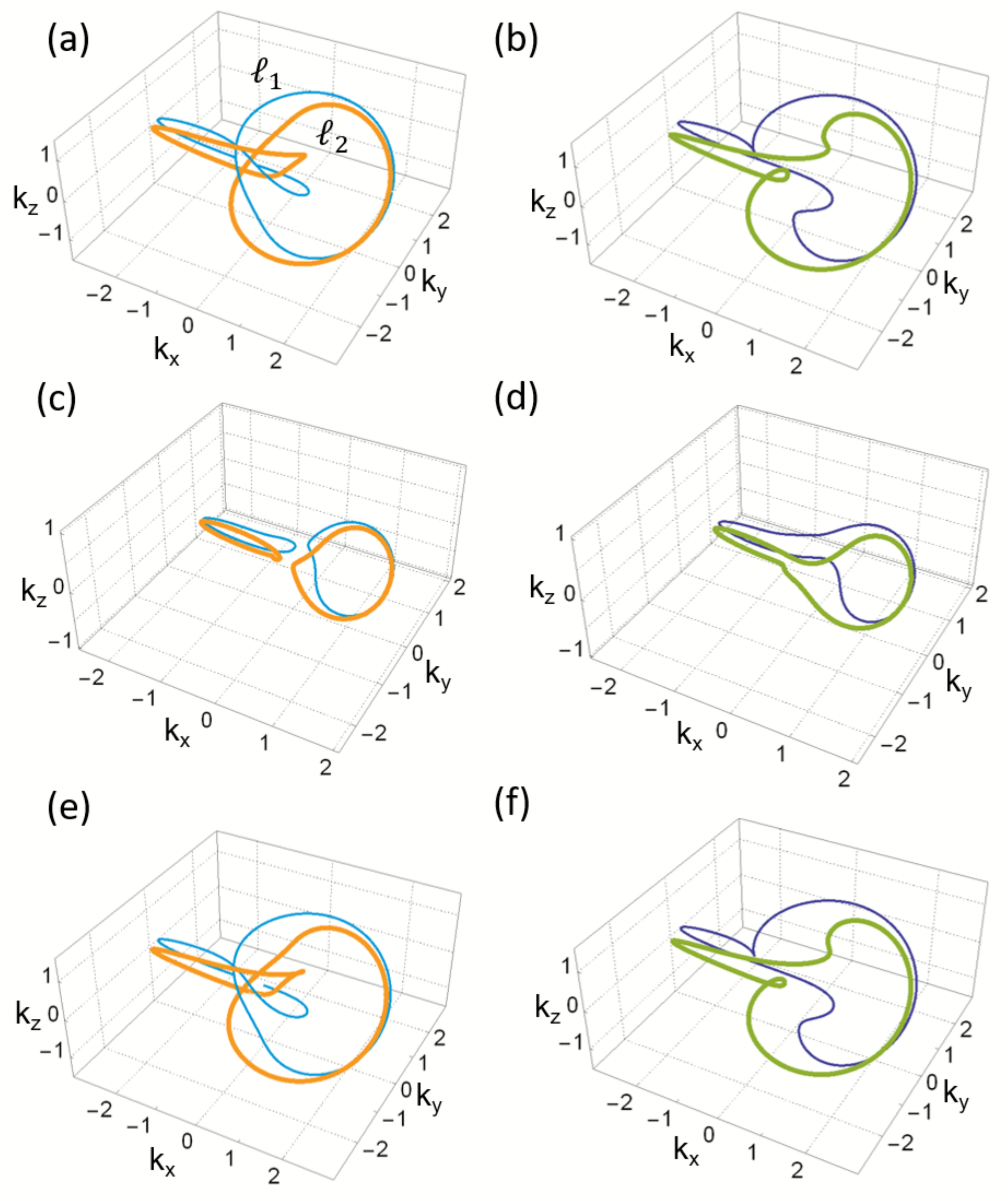}
	\caption{(a)-(d) The exceptional degeneracies in the energy spectra of the model described by Eq. (\ref{eq_ham}) with (a) $m=1$, $\gamma_z=0.8$; (b) $m=1$, $\gamma_z=1.2$; (c) $m=-0.25$, $\gamma_z=0.15$; (d) $m=-0.25$, $\gamma_z=0.3$. (e)-(f) The exceptional degeneracies with an additional symmetry-breaking non-Hermitian perturbation $i\gamma_y\sigma_y$ term; parameters are adopted for (e) $m=1$, $\gamma_z=0.8$, $\gamma_y=0.5$ and (f) $m=1$, $\gamma_z=0.8$, $\gamma_y=1$. $\ell_1$ and $\ell_2$ are two solutions given by Eq. (\ref{eq_es}).}
	\label{fig1}
\end{figure}

For Hermitian systems, a typical two-band nodal-link semimetal can be described by the following 3D continuum model Hamiltonian   \cite{YanNLS2017}
\begin{equation}
\begin{aligned}
H_0(\mathbf{k}) &= [2 k_{x} k_{z}+2 k_{y} (m-k^{2} / 2 ) ] \sigma_{x} \\
&+ [k_{x}^{2}+k_{y}^{2}-k_{z}^{2}- (m-k^{2} / 2 )^{2} ] \sigma_{z}\,,
\end{aligned}\label{eq_hmh}
\end{equation}
where $k^2=\sum_{i=x,y,z} k_i^2$, $\sigma_{x,z}$ are the Pauli matrices and the energy unit is set to be 1. For $m>0$, the band-touching events in the spectrum produce a pair of linked nodal rings, and the link opens when $m<0$.

In this paper, we generalize this nodal-link semimental model to the non-Hermitian case. In the presence of a non-Hermitian term $i\gamma_z\sigma_z$ ($\gamma_z>0$) associated with particle gain and loss for the two spins, the Hamiltonian becomes $H(\mathbf{k})=\sum_{\nu=x,y} h_\nu\sigma_\nu$, where
\begin{equation}
\begin{aligned}
h_x &= [2 k_{x} k_{z}+2 k_{y} (m-k^{2} / 2 ) ] \\
h_z &= [k_{x}^{2}+k_{y}^{2}-k_{z}^{2}- (m-k^{2} / 2 )^{2} ]+i\gamma_z\,.
\end{aligned}\label{eq_ham}
\end{equation}
The eigenenergies are $E_{\pm}=\pm \sqrt{h_x^2+h_z^2}$, which are generally complex for a nonzero $\gamma_z$. Diagonalizing the Hamiltonian (\ref{eq_ham}) leads to a multivalued  characteristic polynomial  $\operatorname{det}[E-H(\boldsymbol{k})]=\prod_{i=1}^2[E-E_{i}(\boldsymbol{k})]$. An exceptional point appearing at $\mathbf{k}_d$ requires that the discriminant of the characteristic polynomial $\operatorname{Disc}_{E}[H](\boldsymbol{k}_d)=0$, with
\begin{equation}
\label{Disc}
\operatorname{Disc}_{E}[\mathcal{H}](\boldsymbol{k})=\prod_{i<j}\left[E_{i}(\boldsymbol{k})-E_{j}(\boldsymbol{k})\right]^{2}\,,
\end{equation}
which implies that $\Re \left(\operatorname{Disc}_{E}[\mathcal{H}](\boldsymbol{k})\right)=\Im \left(\operatorname{Disc}_{E}[\mathcal{H}](\boldsymbol{k})\right)=0$. By solving Eq. (\ref{Disc}) we find that the exceptional points come in pairs and are given by,
\begin{equation}
2 k_y (m-\frac{1}{2} (k_x^2+k_y^2+k_z^2))+2 k_x k_z=\pm \gamma_z\,.\label{eq_es}
\end{equation}
Figure (\ref{fig1}) illustrates the exceptional degeneracies in the spectrum.  For $\gamma_z<m$ [see Fig. \ref{fig1}(a)], the exceptional degeneracies comprise a pair of links, which are reminiscent of a torus link $\mathcal{T}(p,q)$ (double Hopf link) in the general link theory with the node indices $p,q=2$ \cite{Ezawa2017}; for $\gamma_z>m$ [see Fig. \ref{fig1}(b)], the topological structure of the exceptional degeneracies changes and only a pair of unlinked exceptional rings are observed. If $m<0$, the two rings (unlinked in the Hermitian limit) are separated into a tetrad of disconnected exceptional rings for $\gamma_z<|m|$ and form a unlinked pair for $\gamma_z>|m|$, as shown in Figs. \ref{fig1}(c) and \ref{fig1}(d).

The Hamiltonian in Eq. (\ref{eq_ham}) has a chiral symmetry $\mathcal{C} H(\mathbf{k})\mathcal{C}^{-1} =-H(\mathbf{k})$ with $\mathcal{C}\equiv\sigma_{y}$. However, in sharp contrast to the nodal rings (lines) in Hermitian systems that are protected by certain symmetries, the structured exceptional rings do not rely on the symmetries. For instance, we consider a non-Hermitian perturbation with the form of $H_{\rm{NH}}=i\gamma_y\sigma_y+i\gamma_z\sigma_z$, and the exceptional degeneracies in the spectra are illustrated in Figs. \ref{fig1} (e) and \ref{fig1}(d). We can see that the double exceptional Hopf links are robust against non-Hermitian perturbations breaking the chiral symmetry. This can be understood by the dimensionality of the characteristic equation. The exceptional degeneracies are only determined by two constraints $\Re \left(\operatorname{Disc}_{E}[\mathcal{H}](\boldsymbol{k})\right)=0$ and $\Im \left(\operatorname{Disc}_{E}[\mathcal{H}](\boldsymbol{k})\right)=0$. Thus the co-dimension of the Hamiltonian (\ref{eq_ham}) is $2$. One only needs to tune two parameters to have a one-dimensional (1D) gapless nodes in 3D momentum space.

Here we focus on the Hamiltonian in Eq. (\ref{eq_ham}) with the chiral symmetry for simplicity. In this case, we can define a topological charge from the Hamiltonian \cite{Beri2010},
\begin{equation}
Q=\frac{-1}{4 \pi i} \oint_{\mathcal{L}} \operatorname{Tr} [\mathcal{C} (H(\mathbf{k}))^{-1} \nabla_{l} H(\mathbf{k})] d l\,,\label{eq_tq}
\end{equation}
where the integral path $\mathcal{L}$ is around the Fermi surface. If $\mathcal{L}$ encircles the first Brillouin zone, the net topological charge is zero, which is reminiscent of the Fermion doubling or quadrupling problem or the Nielsen-Ninomiya no-go theorem asserting that there is no net chirality in a lattice model of fermions \cite{Nielson1981,Yang2019}. From the modern viewpoint of quantum anomaly, the doubling of Dirac fermions is attributed to the introduction of the lattice \cite{Fujikawa2004,WBRui2018}. In lattice gauge theory, the Wilson fermions show up with a mass associated with the lattice spacing, analogous to the mass term in Pauli-Villars regularization. The Dirac fermions must come in pairs to ensure the theory is anomaly free after the regularization of the latticized theory. However, here the doubling or quadrupling of exceptional degeneracies appears in a continuum model only describing the low-energy sector. This can be understood by introducing cylindrical coordinates $\{k_\rho,\phi,k_z\}$ and decomposing the 3D system into a set of 2D subsystems. Each subsystems exhibits two or four exceptional points. Pictorially, $\Re \left(\operatorname{Disc}_{E}[\mathcal{H}](\boldsymbol{k})\right)=0$ forms disconnected closed paths in each subsystems, and the intersection with the paths defined by $\Im \left(\operatorname{Disc}_{E}[\mathcal{H}](\boldsymbol{k})\right)=0$ always gives exceptional degeneracies of even numbers.

The topological charge given by Eq. (\ref{eq_tq}) is equivalently a winding number $Q=\frac{1}{2\pi}\oint \epsilon_{ij}\tilde{h}_i\partial_l\tilde{h}_j$, where $\tilde{h}_{i(j)}=h_{i(j)}/\sqrt{h_x^2+h_z^2}$ with $i=x,z$. By treating $k_x$ and $k_y$ as parameters, the winding number can be defined for every 1D chain along the $k_z$ direction as
\begin{equation}
w(k_x,k_y)=\frac{1}{2 \pi} \int_{-\infty}^{\infty} d k_{z} \partial_{k_{z}} \phi\,,\label{eq_wn}
\end{equation}
where $\phi \equiv \arctan \left(h_{x} / h_{z}\right)$. One could compact the integral path into a loop by taking equivalence $\phi(-\infty)=\phi(+\infty)$. The integral path is topologically equivalent to an $\mathcal{S}^1$ loop interlinked with the exceptional rings when $k_x$ and $k_y$ lie inside the Fermi surface.

\begin{figure}[htbp]
	\centering
	\includegraphics[width=\textwidth]{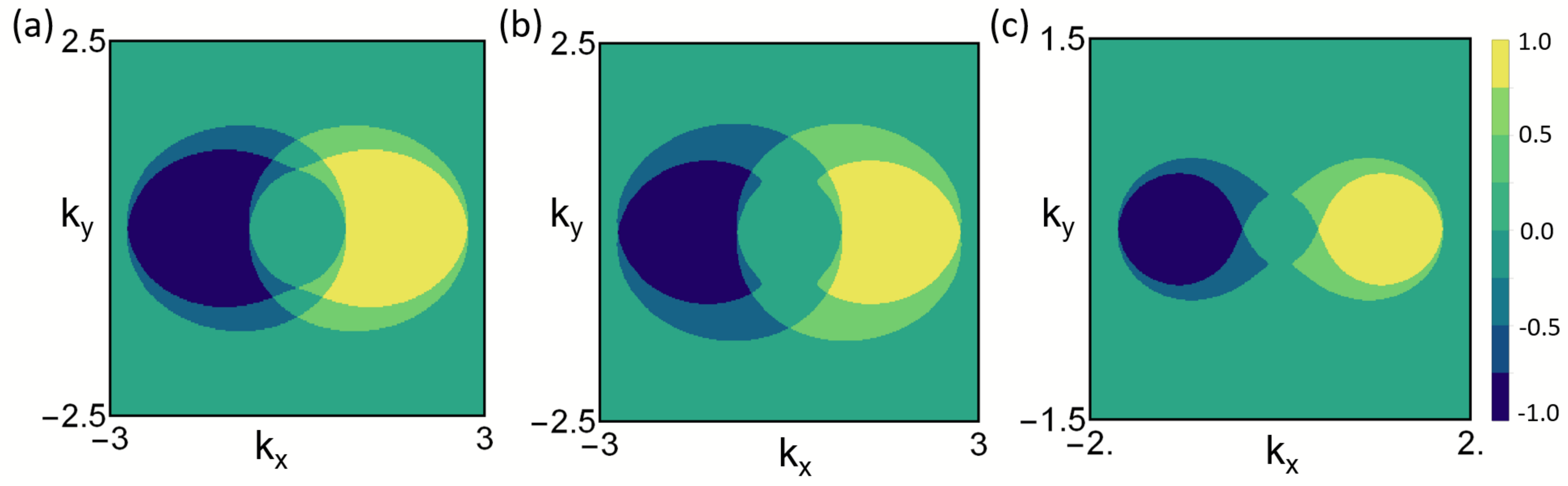}
	\caption{The winding numbers $w(k_x,k_y)$ for (a) $m=1$, $\gamma_z=0.8$; (b) $m=1$, $\gamma_z=1.2$; and (c) $m=-0.25$,$\gamma_z=0.15$. }
	\label{fig2}
\end{figure}

Without loss of generality, we take $m=1$. It is easy to see that $\lim_{k_z \to \infty} \phi=0$. When $k_\rho^4-8 k_\rho^2-4<0$, $\phi$ is discontinuous at
\begin{equation}
k_c^{p,n}=\pm \sqrt{2 \sqrt{2 k_\rho^2-1}-k_\rho^2}\,,
\end{equation}
where we have denoted $k_\rho=k_x^2+k_y^2$. Then the winding number is calculated as  (see Appendix \ref{app_b} for the detailed derivations)
\begin{equation}
w=\frac{1}{2 \pi} \int_{-\infty}^{k_c^n}+\int_{k_c^n}^{k_c^p}+\int_{k_c^p}^{\infty} \partial_{k_z}\phi dk_z\,.
\end{equation}
In contrast, when $k_\rho^4-8 k_\rho^2-4>0$, $\phi$ is continuous and thus gives a trivial winding number.

Figure \ref{fig2} shows the winding numbers $w$ for three typical cases. The winding number takes a multiple of $\mathbb{Z}/2$. The two branches in the exceptional solutions given by Eq. (\ref{eq_es}) carry opposite topological charges. Thus for $k_\rho$ lying around the center of the exceptional degeneracies, the winding number is $0$ since the integral path equivalently interlinks the structure entirely. And when $k_\rho$ lies between the inner and outer exceptional degeneracies, the path encircles only one exceptional point and thus leads to a half-integer value $\pm 1/2$. Furthermore, the $w$ phase diagram distinguishes the double exceptional links by the transition of the half-integer topological charges, as illustrated in Fig. \ref{fig2} (a). The winding number can be associated with a quantized Berry phase,
\begin{equation}
\gamma_B=\oint_{2\mathcal{L}} i\langle \tilde{u}(\mathbf{k})|\partial_{\mathbf{k}}u(\mathbf{k})\rangle d\mathbf{k}\,,
\end{equation}
 via the relation $\gamma_B=2\pi w$, where $\langle\tilde u(\mathbf{k})|$ and $|u(\mathbf{k})\rangle$ are left and right eigenvectors of the Hamiltonian Eq. (\ref{eq_ham}) and the path $2\mathcal{L}$ forms a closed loop on the Riemann surface defined by the complex energy $E_\pm$.

\section{Lattice model and surface states}\label{sec3}

\subsection{Topological zero-energy surface modes}
\begin{figure}[htbp]
	\centering
	\includegraphics[width=\textwidth]{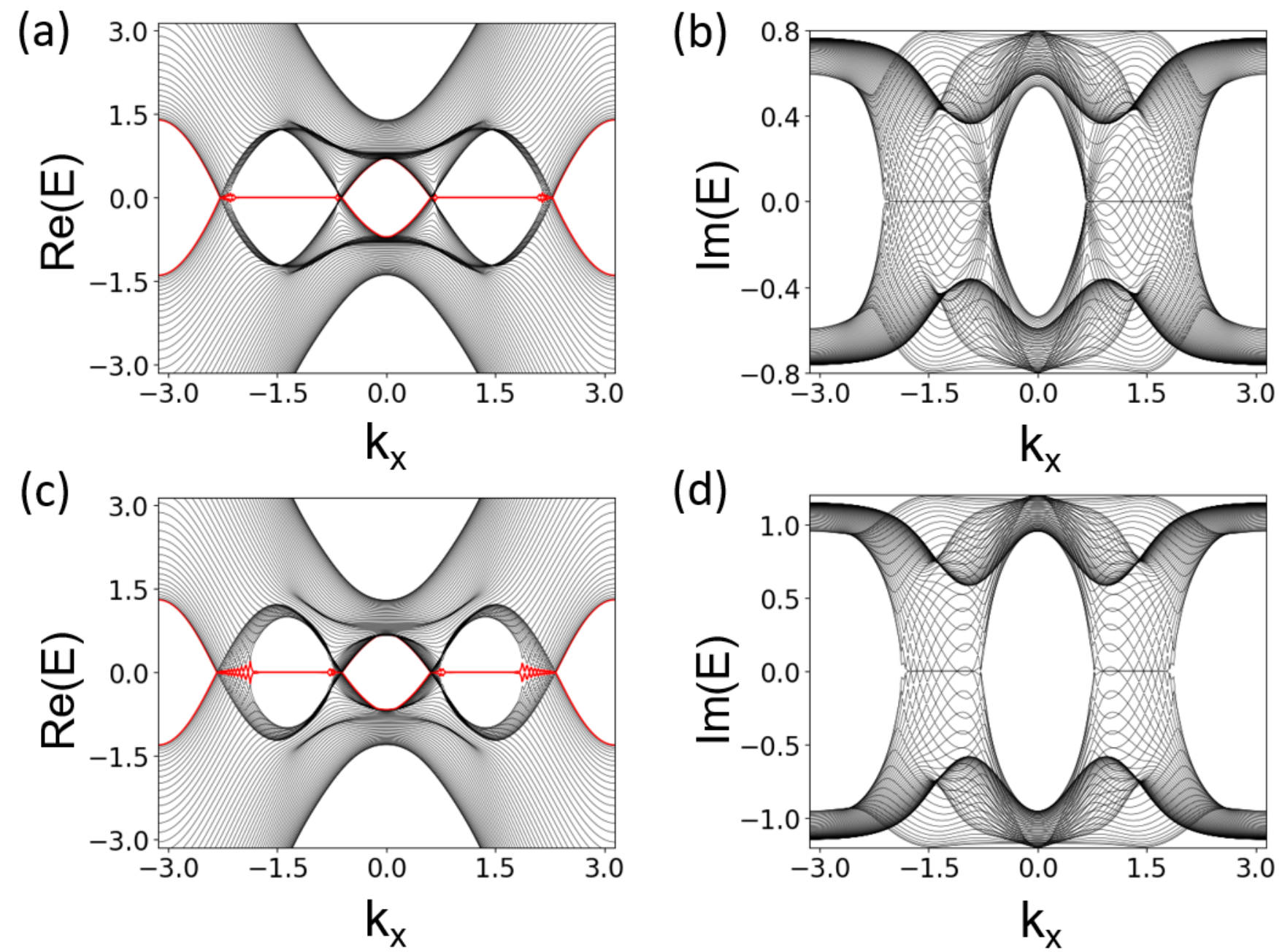}
	\caption{The energy spectra for (a,b) $m=1$, $\gamma=0.8$, $k_y=0.5$ and (c,d) $m=1$, $\gamma=1.2$, $k_y=0.5$ when open boundary condition is imposed along the $z$ direction. The energy unit is set to be 1}
	\label{fig3}
\end{figure}

\begin{figure*}[htbp]
	\centering
	\includegraphics[width=0.9\textwidth]{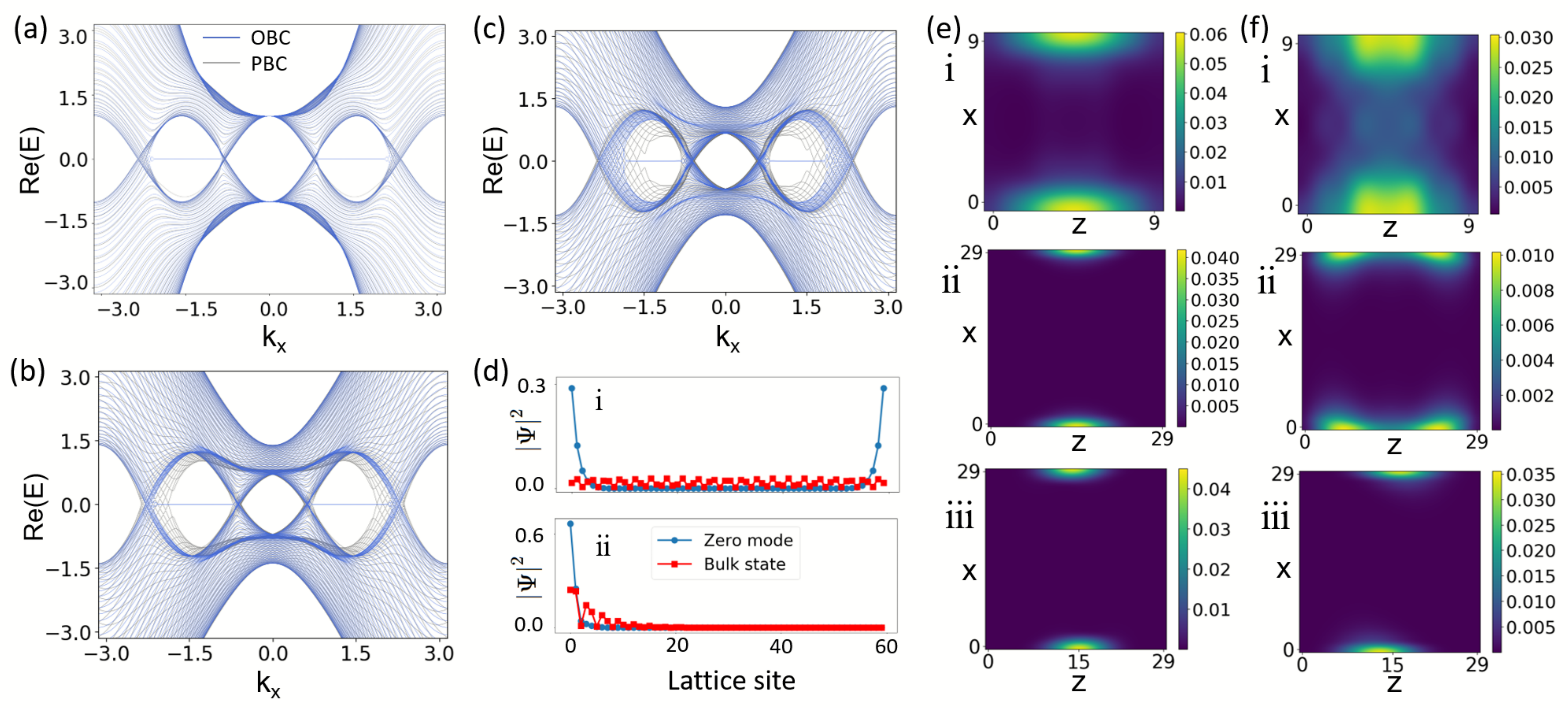}
	\caption{(a)-(c) The energy spectra for (a) $m=1$, $\gamma=0.8$, $k_y=0$,  (b) $m=1$, $\gamma=0.8$, $k_y=0.5$, and (c) $m=1$, $\gamma=1.2$, $k_y=0.5$ when periodic and open boundary condition is imposed along the $z$ direction. 
 (d) The wavefunction distribution for both the zero mode and bulk state for systems with parameters in (a) and (b). (e) The wave function distribution for topological zero modes in the Hermitian limit $\gamma_z=0$ when open boundary condition is imposed along both $x$ and $z$ directions, and for (i) $m=1$, $k_y=0$, and (ii) $m=1$, $k_y=0$, (iii) $m=1$, $k_y=0.5$. (f) The wave function distribution for topological zero modes (and bulk state) corresponding to (e)  for $\gamma_z=1.2$ (when $k_y=0$, there is no skin effect).}
	\label{fig4}
\end{figure*}

In this section, we study the lattice model corresponding to the Hamiltonian in Eq. (\ref{eq_hmh}). To this end, one can construct
 a 3D tight-binding Hamiltonian on a cubic lattice  \cite{YanNLS2017} which is effectively described by the low-energy Hamiltonian (\ref{eq_hmh}), that is,
\begin{widetext}
\begin{equation}
\begin{aligned}
\hat{H}_L=&\sum_{\mathbf{r}}[\hat{H}_{\mathbf{r x}}+\hat{H}_{\mathbf{r y}}+\hat{H}_{\mathbf{r z}}+\hat{H}_{\mathbf{M}}+
\hat{H}_{\mathbf{r x y}}+\hat{H}_{\mathbf{r y z}}+\hat{H}_{\mathbf{r x z}}]\,,~ \hat{H}_{\mathbf{r x}}=m_0 c_{\mathbf{r}}^\dagger \tau_z c_{\mathbf{r}+\mathbf{x}}-\frac{1}{2} c_{\mathbf{r}}^\dagger \tau_z c_{\mathbf{r}+2\mathbf{x}}+\rm{h.c.}\,,\\
\hat{H}_{\mathbf{r y}}=&m_0 (c_{\mathbf{r}}^\dagger \tau_z c_{\mathbf{r}+\mathbf{y}}+c_{\mathbf{r}}^\dagger i\tau_x c_{\mathbf{r}+\mathbf{y}})-\frac{1}{2} (c_{\mathbf{r}}^\dagger \tau_z c_{\mathbf{r}+2\mathbf{y}}+c_{\mathbf{r}}^\dagger i\tau_x c_{\mathbf{r}+2\mathbf{y}})+\rm{h.c.}\,,~
\hat{H}_{\mathbf{r z}}+\hat{H}_{\mathbf{M}}=m_0 c_{\mathbf{r}}^\dagger \tau_z c_{\mathbf{r}+\mathbf{z}}+(m_0^2+1)c_{\mathbf{r}}^\dagger \tau_z c_{\mathbf{r}}+\rm{h.c.}\,,\\
\hat{H}_{\mathbf{r x y}}=&-\frac{1}{2}(c_{\mathbf{r}}^\dagger \tau_z c_{\mathbf{r}+(\mathbf{x}+\mathbf{y})}+c_{\mathbf{r}}^\dagger \tau_z c_{\mathbf{r}+(\mathbf{x}-\mathbf{y})})+ \frac{1}{2}(c_{\mathbf{r}}^\dagger i\tau_x c_{\mathbf{r}+(\mathbf{x}-\mathbf{y})}+c_{\mathbf{r}}^\dagger i\tau_x c_{\mathbf{r}-(\mathbf{x}-\mathbf{y})})+\rm{h.c.}\\
\hat{H}_{\mathbf{r x z}}=&-\frac{1}{2}(c_{\mathbf{r}}^\dagger \tau_z c_{\mathbf{r}+(\mathbf{x}+\mathbf{z})}+c_{\mathbf{r}}^\dagger \tau_z c_{\mathbf{r}+(\mathbf{x}-\mathbf{z})})+
\frac{1}{2}(c_{\mathbf{r}}^\dagger \tau_x c_{\mathbf{r}+(\mathbf{x}-\mathbf{z})}+c_{\mathbf{r}}^\dagger \tau_x c_{\mathbf{r}+(\mathbf{x}+\mathbf{z})})+\rm{h.c.}\,,\\
\hat{H}_{\mathbf{r y z}}=&-\frac{1}{2}(c_{\mathbf{r}}^\dagger \tau_z c_{\mathbf{r}+(\mathbf{y}+\mathbf{z})}+c_{\mathbf{r}}^\dagger \tau_z c_{\mathbf{r}+(\mathbf{y}-\mathbf{z})})+
\frac{1}{2}(c_{\mathbf{r}}^\dagger i\tau_x c_{\mathbf{r}+(-\mathbf{y}+\mathbf{z})}-c_{\mathbf{r}}^\dagger i\tau_x c_{\mathbf{r}+(\mathbf{y}+\mathbf{z})})+\rm{h.c.}\,,
\label{eq_tb}
\end{aligned}
\end{equation}
\end{widetext}
where $c_{\mathbf{r}}=(c_{\mathbf{r},\uparrow},c_{\mathbf{r},\downarrow})^{\rm{T}}$ is the annihilation  operator on site $\mathbf{r}$. The non-Hermiticity is introduced by an on-site gain and loss $H_N=\sum_\mathbf{r}( i\gamma_z c_{\mathbf{r},\uparrow}^\dagger c_{\mathbf{r},\uparrow}-i\gamma_zc_{\mathbf{r},\downarrow}^\dagger c_{\mathbf{r},\downarrow})$. Under periodic boundary condition and with Fourier transformation, the tight-binding Hamiltonian can be written as $\hat{H}_{\mathbf{k}}=\sum_{\mathbf{k}, s s^{\prime}} c_{\mathbf{k} s}^{\dagger}[\mathcal{H}(\mathbf{k})]_{s s^{\prime}} c_{\mathbf{k} s^{\prime}}$, where $c_{\boldsymbol{k} s}=1 / \sqrt{V} \sum_{\boldsymbol{r}} e^{-i \boldsymbol{k} \cdot \boldsymbol{r}} c_{\boldsymbol{r} s}$ is the annihilation operator in momentum space and the Bloch Hamiltonian is written as
\begin{equation}
\mathcal{H}(\boldsymbol{k})=d_x(\boldsymbol{k})\sigma_x+d_z(\boldsymbol{k})\sigma_z,\label{eq_latt}
\end{equation}
where
\begin{equation*}
\begin{split}
d_x&=2 \sin k_{x} \sin k_{z}+2 \sin k_{y}(\sum_{i=x, y, z} \cos k_{i}-m_{0})\,,\\
d_{z}&=\sin ^{2} k_{x}\!+\! \sin ^{2} k_{y}\!-\!\sin ^{2} k_{z}\!-\!(\sum_{i=x, y, z}\! \cos k_{i}-m_{0})^{2}\!+\!i\gamma_z.
\end{split}
\end{equation*}
We numerically calculated the energy spectra $E(k_x)$ of a reduced chain with length $L_z=60$ and fixed $k_y$ under open boundary condition (OBC) along the $z$ axis. The results are shown in Fig. \ref{fig3}. In the presence of non-Hermiticity, the degeneracy of the surface states is lifted near the band-touching points of the bulk bands. The crescent-like flat surface states connecting the two links are draped on the edge, which implies the doubling structure of the exceptional points. In fact, the lattice Hamiltonian in Eq. (\ref{eq_tb}) has anomalous bulk-boundary correspondence, with typical energy spectra under periodic and open boundary conditions shown in Figs. \ref{fig4} (a)-(c). The discrepancy between the PBC and OBC  spectra for $k_y \neq 0$ signifies the breakdown of the bulk-boundary correspondence. When the conventional bulk-boundary correspondence breaks down, the bulk states localize near the edge, which is dubbed as the non-Hermitian skin effect \cite{ZWang2018,FSong2019,LiGong2019,CHLee2019,LeeHybrid2019,Hofmann2019,OkumaTopo2019,CHLeeUn2019}, as shown in Fig. \ref{fig4} (d). A non-Bloch theory is built via complex analytical continuation of the Bloch momentum $k \to k+i\kappa$ \cite{ZWang2018} to restore the bulk-boundary correspondence of this system; see Sec. \ref{sec_gbz} for details.

Notably, we find that this non-Hermitian system has intriguing size-dependent topological zero-energy surface modes. We numerically calculate the eigen-wave-function distributions closest to the zero energy, shown in Figs. \ref{fig4} (e) and \ref{fig4}(f). When the open boundary condition is imposed along both $x$ and $z$ directions, the topological surface modes depend on the system size. In the Hermitian limit, the surface states are squeezed toward two sides of the lattice. In the presence of a large non-Hermiticity, $e.g.$, $\gamma_z=1.2$, the topological surface modes are gapped out for small lattice size, as illustrated in Fig. \ref{fig3} (f) i. Upon the size increasing to $30\times 30$ (in the $x$-$z$ plane), the topological zero modes survive but with the localized center drifting as $k_y$ changes. The size-dependent topological phase transition was revealed in a 1D Hermitian \cite{ZhaoT2017} and non-Hermitian systems \cite{LLiC2020}. We here show that this phase transition can be generalized to a higher dimensional system.

To give an intuitive insight on this critical phenomena, we connect the boundaries of the lattice on both $x$ and $z$ directions, which gives rise to an effective 1D description. By treating $k_x$ and $k_y$ as parameters, the Hamiltonian  (\ref{eq_latt}) effectively reduces to a 1D Hamiltonian given by
\begin{equation}
\begin{split}
H_{\mathrm{eff}}=&[t_x \sin k_z+t_y(\cos k_z+m_{xy})]\sigma_x\\&+(-2m_{xy}\cos k_z+M+i\gamma_z)\sigma_z\,,\label{eq_1D}
\end{split}
\end{equation}
where $t_x\equiv 2\sin k_x$, $t_y\equiv 2\sin k_y$, $m_{xy}\equiv \cos k_x+\cos k_y-m_0$, and $M=\sin^2 k_x+\sin^2 k_y-m_{xy}^2-1$. For simplicity, we focus on a shortcut solution for $k_y=0$. With a basis change $\sigma_z \to \sigma_y$, the Hamiltonian Eq. (\ref{eq_1D}) bears a resemblance to a $\pi$-flux Su-Schrieffer-Heeger model, $H_{S}=t_1\sin(k) \sigma_x+(t_2\cos(k)+t_3+i\gamma)\sigma_y$. In real space, we take a similarity transformation $\tilde{H}_S=S^{-1}H_S S$ with a diagonal matrix $S=\mathrm{diag}(1,r,r,r^2,r^3,r^4,r^6,r^7\ldots)$, where we have chosen a basis $|\psi\rangle=(\psi_{1,A},\psi_{1,B},\psi_{2,A},\psi_{2,B},\ldots,\psi_{L,A},\psi_{L,B})^T$. The transformed Hamiltonian reads
\begin{equation}
\begin{split}
\tilde{H}_S=&\sum_{i} \tilde{t}c^\dagger_{1,A}c_{i,B}+\tilde{t}^*c^\dagger_{i,B}c_{1,A}\\
&+i\tilde{t}_2r^{i+1}c_{i,A}^\dagger c_{i+1,B}-\frac{i\tilde{t}_2}{r^{i+1}}c_{i+1,B}^\dagger c_{i,A}\\
&-i\tilde{t}_1 r^{i-1} c_{i,B}^\dagger c_{i+1,A}+\frac{i\tilde{t}_1}{r^{i-1}} c_{i+1,B}^\dagger c_{i,B}\,,
\end{split}
\end{equation}
where we have chosen $r=\sqrt{(it_3-\gamma)/(\gamma-it_3)}$ and redefined $\tilde{t}_1=\frac{t_1-t_2}{2}$ and $\tilde{t}_2=-\frac{t_1+t_2}{2}$. The non-Hermiticity is transfered from intracell hopping to intercell hopping, with a site-dependent amplitude $\sim r^L$. The strength of the non-reciprocity will dominate with increasing magnitude when going deep into the bulk, which would account for the critical behavior depending on the lattice size.

\subsection{Non-Bloch theory for the bulk-boundary correspondence}\label{sec_gbz}

Now we present a non-Bloch theory to restore the bulk-boundary correspondence in this 3D non-Hermitian system. We take an ansatz for the real-space wave function as a linear combination \cite{ZWang2018,Murakami2019},
\begin{equation}
\psi_{n, \mu}=\sum_{j} \phi_{n, \mu}^{(j)},\quad \phi_{n, \mu}^{(j)}=\left(\beta_{j}\right)^{n} \phi_{\mu}^{(j)}\,,
\end{equation}
where $\mu=\mathrm{A}, \mathrm{B}$ denotes the two spins. By imposing that the $\phi_{n, \mu}^{(j)}$ is an eigenstate of $\mathrm{det}[H_{\rm{eff}}-E]=0$, one can obtain
\begin{equation}
\begin{split}
&[-\frac{t_x}{2}i(\beta-\beta^{-1})+\frac{t_y}{2}(\beta+\beta^{-1})+t_y m_{xy}]\phi_B\\&+[-m_{xy}(\beta+\beta^{-1})+M+i\gamma_z]\phi_A=E \phi_A\,,\\
&[-\frac{t_x}{2}i(\beta-\beta^{-1})+\frac{t_y}{2}(\beta+\beta^{-1})+t_y m_{xy}]\phi_A\\&-[-m_{xy}(\beta+\beta^{-1})+M+i\gamma_z]\phi_B=E \phi_B\,.
\end{split}\label{eq_gbz}
\end{equation}
Thus the generalized Bloch Hamiltonian $\mathcal{H}(\beta)$ can be written as
\begin{equation}
\mathcal{H}(\beta)=R_x(\beta)\sigma_x+R_z(\beta)\sigma_z,
\end{equation}
where
\begin{equation*}
\begin{split}
& R_x(\beta)=-\frac{t_x}{2}i(\beta-\beta^{-1})+\frac{t_y}{2}(\beta+\beta^{-1})+t_y m_{xy},\\
& R_z(\beta)=-m_{xy}(\beta+\beta^{-1})+M+i\gamma_z.
\end{split}
\end{equation*}

\begin{figure}[htbp]
	\centering
	\includegraphics[width=\textwidth]{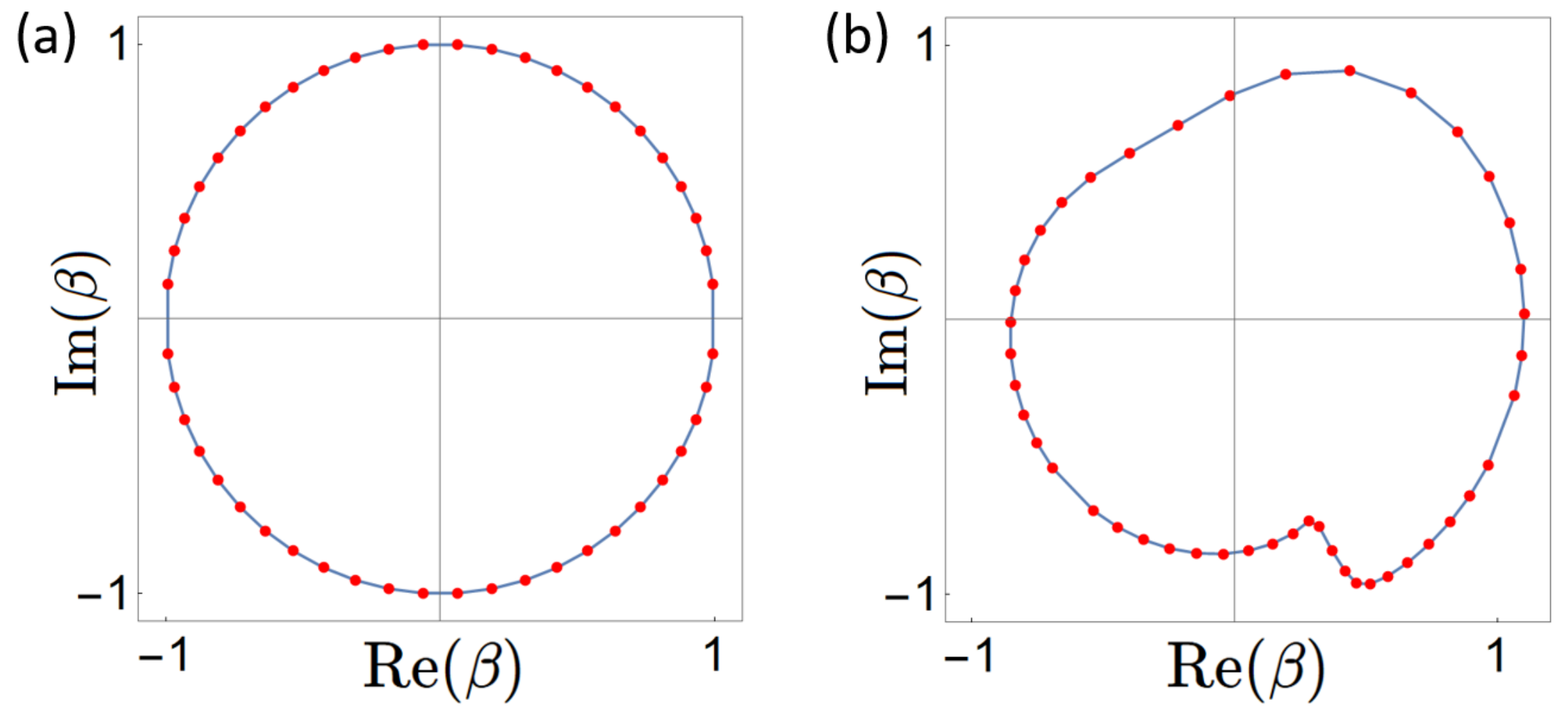}
	\caption{The generalized Brillouin zone for (a) $k_x=\pi/4$, $k_y=0$, $\gamma_z=0.8$, $m_0=1$ and (b) $k_x=\pi/4$, $k_y=0.5$, $\gamma_z=0.8$, $m_0=1$.}
	\label{fig5}
\end{figure}

In general, Eq. (\ref{eq_gbz}) has four solutions, which can be labeled as $\beta_s$ ($s=1,2,3,4$). The generalized Brillouin zone (GBZ) can be determined from the two solutions $|\beta_2|=|\beta_3|$ by requiring the continuum conditions $\beta_1\le|\beta_2|=|\beta_3|\le |\beta_4|$. We numerically calculate the GBZ for our model. The results are shown in Fig. \ref{fig5}. For $k_y=0$, the ordinary bulk-boundary correspondence holds such that the GBZ constitutes a unit circle (since $|\beta|=1$), as illustrated in Fig. \ref{fig5} (a). When $|\beta|\neq 1$, the skin modes in bulk states appear and the bulk-boundary correspondence breaks down.

Following the method outlined in Ref. \cite{ZWang2018}, we introduce a non-Bloch winding number to characterize the open-bulk spectra, that is,
\begin{equation}
\chi=\frac{1}{2\pi}\int_{C_\beta}\epsilon_{ij} \tilde{h}_i\partial_l\tilde{h}_j\,,
\end{equation}
where $C_\beta$ denotes the GBZ, and $i=x,z$. The topological zero modes in OBC spectra are protected by a nonzero topological index $\chi$. For instance, $\chi=1$ when $k_x=1.5$, $k_y=0$ and $\gamma_z=0.8$.

\section{Proposed realization and detection in optical lattices}\label{sec4}

\begin{figure}[htbp]
	\centering
	\includegraphics[width=\textwidth]{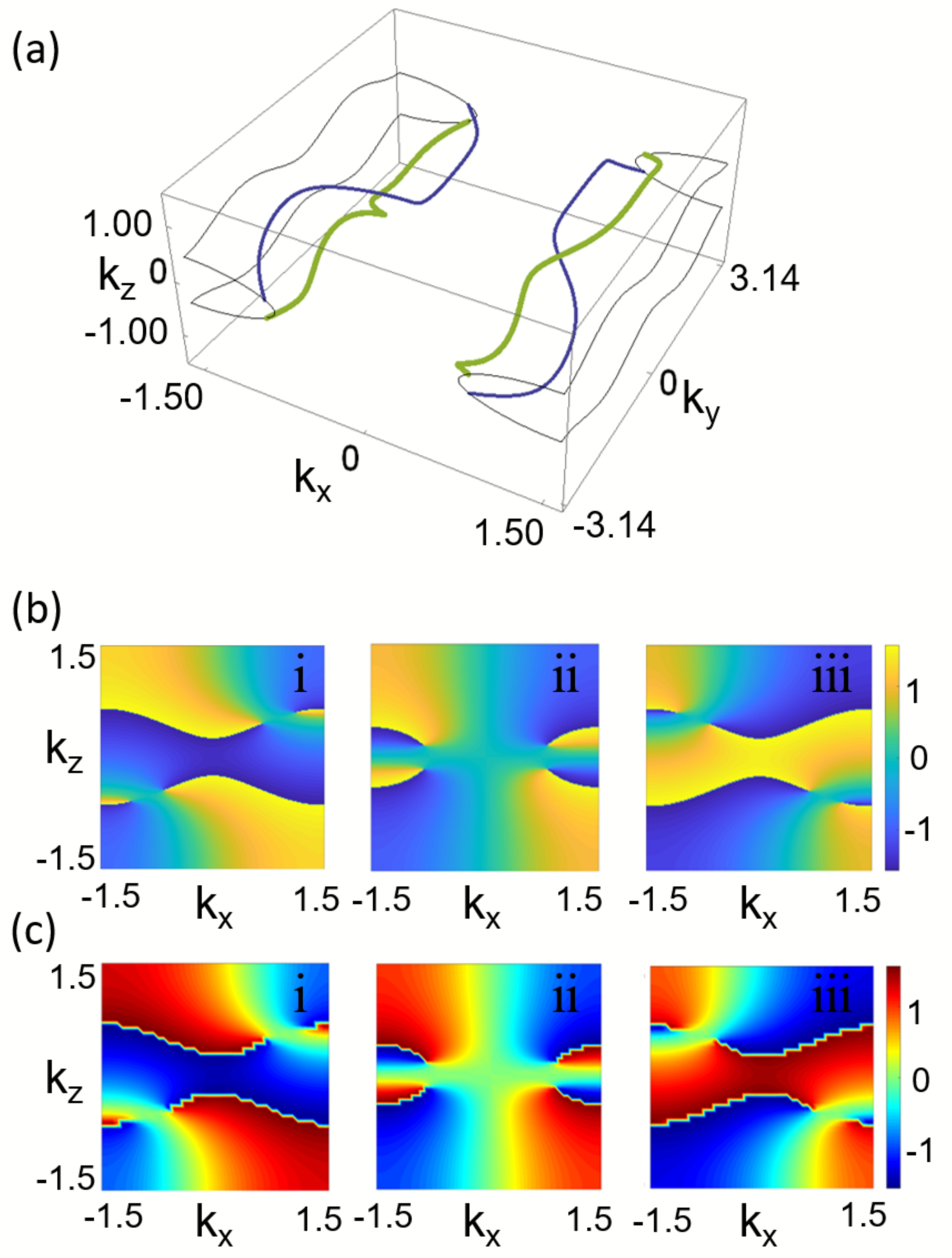}
	\caption{(a) The exceptional nodal rings in the spectrum of Hamiltonian (\ref{eq_apr}) with $m=-0.5$ and $g=0.25$; (b) $\Re(\phi)$ in Eq. \ref{eq_wn} with parameters in (a) and with (i) $k_y=1$, (ii) $k_y=0$ and (iii) $k_y=-1$; (c) $\Re(\eta_{xz})$ defined from long-time averaged spin textures with parameters in (a) and with (i) $k_y=1$, (ii) $k_y=0$, and (iii) $k_y=-1$. The evolution time is $T=80$.}
	\label{fig6}
\end{figure}

In this section, we discuss a possible scheme to realize the double exceptional links with a dissipative cold atomic gas.
The realization of the Hamiltonian in Eq. (\ref{eq_tb}) with cold atoms in optical lattices is challenging due to complicated configuration of the spin-orbit coupling involved in the hopping terms. However, for small $m$, the Hamiltonian in Eq. (\ref{eq_ham}) can be approximated by
\begin{equation}
\begin{split}
\tilde{H}(\mathbf{k})\approx &(2 k_{x} k_{z}+2 m k_{y}) \sigma_x+ [(m+1)k_{x}^{2}\\
&+(m+1)k_{y}^{2}+(m-1)k_{z}^{2}- m^{2} +i\gamma_z] \sigma_z\,,
\end{split}
\end{equation}
which emerges as low-energy excitations of a lattice model
\begin{equation}
\begin{split}
\tilde{\mathcal{H}}(\mathbf{k})\approx &[2 \sin(k_{x}) \sin(k_{z})+2 m \sin(k_{y}) ] \sigma_x+ [t_x \sin^{2}(k_{x})+\\
&t_y \sin^{2}(k_{y})+t_z \sin^{2}(k_{z})- m^{2} +i\gamma_z] \sigma_z.
\end{split}\label{eq_apr}
\end{equation}
Here we denote $t_x \equiv 1+m$, $t_y \equiv 1+m$, and $t_z \equiv m-1$. Two pairs of exceptional lines are observed around the center of the Brillouin zone, as shown in Fig. \ref{fig6}(a). Due to the periodicity of the Brillouin zone, this equivalently forms a Hopf link. Thus the exceptional nodes of this simplified model have structures identical to that of the model (\ref{eq_latt}). The hopping terms required to realize this model can be diagrammatically visualized as,
\begin{equation}
 \begin{aligned}
&T_x=|1\rangle\overset{-\frac{t_x}{4}}{\tilde{\curvearrowleft} }|1\rangle+|2\rangle\overset{\frac{t_x}{4}}{\tilde{\curvearrowleft} }|2\rangle+\text{
H.c.}\,,\\
&T_y=|1\rangle\overset{-\frac{t_y}{4}}{\tilde{\curvearrowleft} }|1\rangle+|2\rangle\overset{\frac{t_y}{4}}{\tilde{\curvearrowleft} }|2\rangle\\
&+|1\rangle\overset{-im}{\curvearrowleft }|2\rangle \overset{im}{\curvearrowright }|1\rangle+\text{
H.c.}\,,\\
&T_z=|1\rangle\overset{-\frac{t_z}{4}}{\tilde{\curvearrowleft} }|1\rangle+|2\rangle\overset{\frac{t_z}{4}}{\tilde{\curvearrowleft} }|2\rangle+\text{
H.c.}\,,\\
&T_{x+z}=|1\rangle\overset{-\frac{1}{2}}{\curvearrowleft }|2\rangle\overset{\frac{1}{2}}{\curvearrowright }|1\rangle+\text{
H.c.}\,,\\
&T_{x-z}=|1\rangle\overset{-\frac{1}{2}}{\curvearrowleft }|2\rangle\overset{\frac{1}{2}}{\curvearrowright }|1\rangle+\text{
H.c.}\,,\\
\end{aligned}\label{eq_hop}
\end{equation}
where $\curvearrowright$ denotes the nearest hopping and $\tilde{\curvearrowright}$ denotes the next-nearest hopping along the corresponding direction. The mass term is given by $T_{M}=M(|1\rangle\langle1|-|2\rangle\langle2|)$.

\begin{figure}[htbp]
	\centering
	\includegraphics[width=\textwidth]{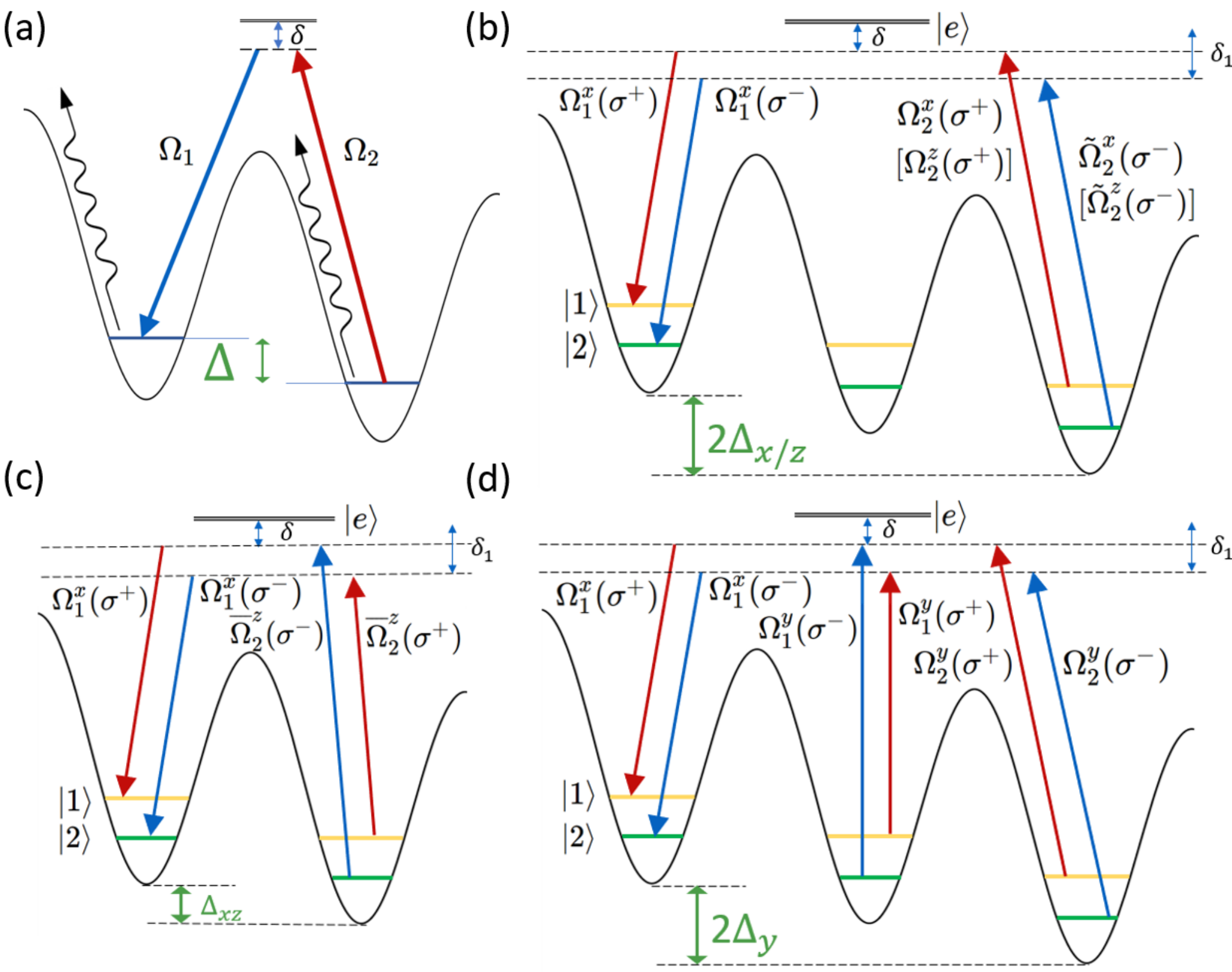}
	\caption{Schematics of the laser configurations to realize the Hamiltonian Eq. (\ref{eq_apr}).  Polarizations of each beam are shown in brackets. Rabi frequencies for each beam are: $\Omega_1^x=\Omega_0 e^{i kx}$, $\Omega_2^{x(z)}=\frac{t_{x(z)}}{4}\Omega_{x(z)} e^{-ikx(z)}$, $\tilde{\Omega}_2^{x(z)}=-\frac{t_{x(z)}}{4}\Omega_0 e^{-ikx(z)}$, $\overline{\Omega}_2^z(\sigma^\pm)=\pm\frac{1}{2}\Omega_{z\pm}e^{ikz}$,$\Omega_1^y(\sigma^\pm)=\pm im\Omega_{y1} e^{iky}$, $\Omega_2^y(\sigma^\pm)=\pm \frac{t_y}{4}\Omega_{y1} e^{iky}$.}
	\label{fig7}
\end{figure}

Two-photon Raman-assisted transitions are utilized to achieve the hopping terms involved in Eq. (\ref{eq_hop}) \cite{STWang2014,DWZhang2015,YQZhu2017}. In experiments, a moderate magnetic field is applied to distinguish the spin states in the ground state manifold. The optical lattice is tilted with a homogeneous energy gradient $\Delta_{x,y,z}$ along the $x$,$y$,$z$ directions to suppress the natural hopping $t_0$ with $t_0 \ll \Delta_{x,y,z}$. And we require $\Delta_x \neq \Delta_y \neq \Delta_z \neq \Delta_x \pm \Delta_z$ to ensure the needed broken parity (left-right) symmetry. The internal states $|1\rangle$, $|2\rangle$ differ in the magnetic quantum number by 1, thus are coupled to the excited state $|e\rangle$ by a ${\sigma}^+$-polarized and ${\sigma}^-$-polarized light respectively. The excited state $|e\rangle$ can be adiabatically eliminated from the Raman transition for a large detuning $\delta$. Between the sites $\mathbf{r}$ and $\mathbf{r} + \mathbf{m}$, the Raman-assisted hopping rate is given by
\begin{equation}
t_{\mathbf{r}, \mathbf{m}}=\frac{\Omega_{\beta \mathbf{m}}^{*} \Omega_{\alpha \mathbf{m}}}{\delta} \int d^{3} \mathbf{r}^{\prime} w^{*}\left(\mathbf{r}^{\prime}-\mathbf{r}-\mathbf{m}\right) e^{i \delta \mathbf{k} \cdot \mathbf{r}^{\prime}} w (\mathbf{r}^{\prime}-\mathbf{r} )\,,
\end{equation}
where $w (\mathbf{r}^{\prime}-\mathbf{r} )$ is the Wannier-Stark function at the site $\mathbf{r}$ and $\delta \mathbf{k}=\mathbf{k}_{\alpha}-\mathbf{k}_{\beta}$ is the momentum difference between the relevant Raman beams with the corresponding single-photon Rabi frequencies $\Omega_{\alpha \mathbf{m}}$ and $\Omega_{\beta \mathbf{m}}$.

In principle, arbitrary hopping matrices including the required hopping terms can be independently created with well-designed laser configurations, so we only take one of the hopping terms in $T_x$ $|1\rangle\overset{-t_x/4}{\tilde{\curvearrowleft} }|1\rangle$ and $|2\rangle\overset{t_x/4}{\tilde{\curvearrowleft} }|2\rangle$ as an example. The relevant Raman pair is chosen as $\Omega_1^x=\Omega_0 e^{i kx}$ ($\sigma^+$ polarized) and $\Omega_2^x=-\frac{t_x}{4}\Omega_0 e^{-ikx}$ ($\sigma^+$ polarized) to introduce the hopping in $|1\rangle$ states. For this case $\delta \mathbf{k}$ takes the value of $(2k,0,0)$.Thus the site dependent phase term can be reduced to $e^{i\delta \mathbf{k}\cdot \mathbf{r}}=1$ by adjusting the interfering angle of the lattice beams so that $ka=2\pi$. Similarly, the Raman pair to generate the hopping $|2\rangle\overset{t_x/4}{\tilde{\curvearrowleft} }|2\rangle$ can be chosen as $\Omega_1^x=\Omega_0 e^{i kx}$ ($\sigma^-$ polarized)) and $\tilde{\Omega}_2^x=\frac{t_x}{4}\Omega_x e^{-ikx}$ ($\sigma^-$ polarized) with $\Omega_x=\Omega_0\delta_1/\delta$. Similarly, all the hopping terms in Eq. (\ref{eq_hop}) can be introduced with lasers beams using configuration illustrated in Fig. \ref{fig7}. To demonstrate that the next-nearest hopping is feasible with current technology, we take $^{40}\mathrm{K}$ atoms of mass $m=66.422\times 10^{-27}~\mathrm{kg}$ in an optical lattice as an example. The optical lattice can be generated by laser beams at wavelength $\lambda=764~\rm{nm}$ with depth $V_0\approx 2.3 E_R$, where $E_R/\hbar=\hbar k_L^2/2m \approx 2\pi\times 8.545~\mathrm{kHz}$ is the recoil energy. The overlap ratio $\Lambda \sim 0.12$ can be estimated by the overlap integral of the Wannier-Stark function (see Appendix \ref{appd} for details).  Notably, the Raman beams must be properly set to adiabatically eliminate the excited-state manifold in the Raman scheme. In experiments, this can be achieved by using the lowest-energy excited state (e.g., the $|P_{1/2}\rangle$ state) as the state $|e\rangle$ in Fig. \ref{fig7} and setting the frequencies of the Raman beams to be large and red detuning from $|e\rangle$. For Raman beams with $\Omega_0/2\pi=30~\mathrm{MHz}$ and the single-photon detuning $\delta/2\pi=1.5~\mathrm{THz}$, one has $\Omega _{0}^{2}/\delta \approx 2\pi \times 0.6\,$kHz and the Raman-assisted hoping rate $t/\hbar\approx2\pi\times 72~\mathrm{Hz}$. The population of the excited state $|e\rangle$ and the effective spontaneous emission rate are both negligible, and are estimated by $\Omega_{0}^2/\delta^{2}$ and $\Gamma _{s}\Omega_{0}^2/\delta^{2}$ with the decay rate of the excited state $\Gamma _{s}\approx 2\pi \times 6\,$MHz, respectively. We note that the proposed scheme could be realized by using other alkaline atoms such as $^{6}$Li and boson species $^{87}$Rb and $^{23}$Na.

To introduce non-Hermiticity into the lattice, on-site atom loss $-2i\gamma_z$ in the $|2\rangle$ state can be generated by applying a radio frequency pulse to resonantly transfer the atoms to an irrelevant state $|\tilde{e}\rangle$ [Fig. \ref{fig5} (a)]. This only differs from the model Hamiltonian (\ref{eq_apr}) by a background gain $e^{-\gamma_z t}$. Notably, this method has been adopted to experimentally realize non-Hermitian Hamiltonians with atomic gases \cite{JLi2019} and superconducting qubits \cite{Naghiloo2019}.

We propose to measure the exceptional Hopf links and the winding number by probing the dynamics of the momentum distribution $\rho(\mathbf{k})$ on each spin state. To this end, we first define the spin textures as the time-dependent expectation values of the Pauli matrices $\langle\sigma_j(\mathbf{k},\tau)\rangle=\langle\tilde{u}_\mathbf{k}|\sigma_j|u_\mathbf{k}\rangle/\langle\tilde{u}_\mathbf{k}|u_\mathbf{k}\rangle$ in a biorthogonal formalism, where $\tau$ denotes the time. The winding number in Eq. (\ref{eq_wn}) can be extracted from the long-time averaged spin vector $(\overline {\sigma}_x,\overline {\sigma}_z)$ \cite{BZhu2019,LZhou2019},
\begin{equation}
w_d=\frac{1}{2\pi}\oint_{\mathcal{L}} \partial_\mathbf{k} \eta_{xz}(\mathbf{k})d\mathbf{k}\,,
\end{equation}
where $\eta_{xz}(\mathbf{k})\equiv \arctan(\overline{\sigma}_x/\overline{\sigma}_z)$ and $\overline{\sigma}_j=\frac{1}{T}\int_0^T\langle\sigma_j(\mathbf{k},\tau)\rangle d\tau$. When the Hamiltonian is non-Hermitian, $\eta_{xz}$ is generally a complex angle. It can be decomposed as the sum of two observables
$$\Re(\eta_{ij}(\mathbf{k}))=\frac{1}{2}(\phi_{ij}^{RR}+\phi_{ij}^{LL}),$$ where $$\phi_{ij}^{RR}=\arctan(\overline{\langle u|\sigma_i|u\rangle}/\overline{\langle u|\sigma_j|u\rangle}),$$ $$\phi_{ij}^{LL}=\arctan(\overline{\langle \tilde{u}|\sigma_i|\tilde{u}\rangle}/\overline{\langle \tilde{u}|\sigma_j|\tilde{u}\rangle}).$$
The numerical results show that for our model $\phi_{ij}^{RR}=\phi_{ij}^{LL}$. Thus merely measuring $\phi^{RR}$ is enough to determine the topology in the experiments. Furthermore, the bulk nodes can be mapped out by reconstructing the $k_y$-resolved spin textures. Due to its topological nature as a defect, the exceptional point is identified by the singularity in $\Re(\eta_{xz})$. As illustrated in Fig. \ref{fig4} (b), the linked structure can be reconstructed by projected singularities with opposite topological polarities. In experiments, the spin texture $\langle \sigma_z\rangle$ can be measured by the quasimomentum distribution $\rho(k_x,k_z)$ on each spin state by the time-of-flight measurement after abruptly turning off the lattice potential, while the spin texture $\langle \sigma_x\rangle$ can be transferred from the spin population difference by a $\pi/2$ pulse. We numerically calculate the long-time averaged $\eta_{xz}$, as shown in Fig. \ref{fig6} (c), which is in good agreement with $\phi$ [defined in Eq. (\ref{eq_wn})].

\begin{figure}[htbp]
	\centering
	\includegraphics[width=\textwidth]{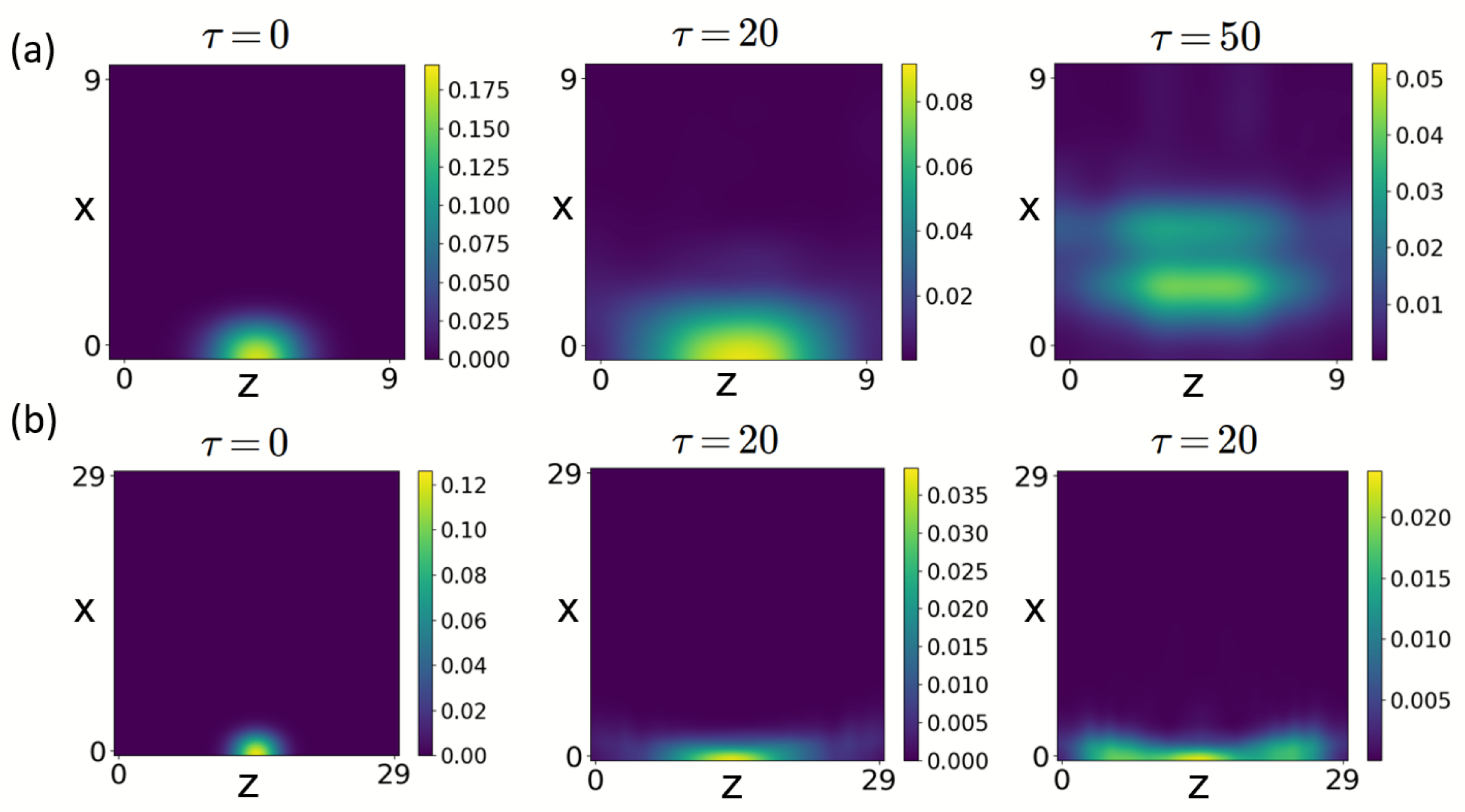}
	\caption{Snapshots of the density distribution at different evolution times (denoted by $\tau$) for (a) a smaller lattice and (b) a bigger lattice. The initial state takes the Gaussian form for (a) $\psi(\tau=0)=\mathcal{N} \exp [-(x-15)^{2} / 10-z^{2} / 5](1,1)^{T}$ and (b) $\psi(\tau=0)=\mathcal{N} \exp [-(x-5)^{2} / 3-z^{2} / 0.5](1,1)^{T}$, where $\mathcal{N}$ is the normalization factor. The intensity profile of an evolved state $\psi(\tau)$ is normalized.}
	\label{fig8}
\end{figure}

Last, we address an experimental approach to detect the localized boundary states. For cold atoms in an optical lattice, one could initially set the atoms in a Gaussian distribution and then observe the particle density evolution. We illustrate the numerically simulated results (via the solution of the Schr\"{o}dinger equation) of detection signatures  in Fig. \ref{fig8}. For a small lattice size $10 \times 10$ (in the $x$-$z$ plane) in the absence of topological zero modes, the wave packet enters the bulk. In contrast, one can see clear signatures of localized edge modes, which have a distribution similar to that of Fig. \ref{fig4} (f), plot ii.

\section{Conclusion}\label{sec5}
In summary, we have investigated the topological properties of an exceptional double Hopf link in a non-Hermitian system. The topology of the exceptional double link is characterized by a winding number or a quantized Berry phase. The topological properties of the corresponding lattice model are addressed and a non-Bloch theory is developed for this system to describe the anomalous bulk-boundary correspondence. We have further proposed a feasible scheme to realize and detect the exceptional Hopf link with dissipative cold atoms in a 3D optical lattice.
The proposed system would provide a promising platform for elaborating the exotic physics of the non-Hermitian topological semimetals.

\acknowledgments

The authors thank Yan-Qing Zhu and Yu-Guo Liu for useful discussions.
The work was supported by  the National Key Research and Development Program of China (Grant No. 2016YFA0301803), the National Natural Science Foundation of China (Grant No. 91636218, No. U1801661, and No. U1830111), the Key-Area Research and Development Program of GuangDong Province (Grant No. 2019B030330001), and the Key Project of Science and Technology of Guangzhou (Grant No. 201804020055).

\begin{appendix}

\section{Derivation of the winding number}\label{app_b}
The winding number for a general two-band model $H=h_x\sigma_x+h_z\sigma_z$ can be decomposed to the wighted sum of two components,
\begin{equation}
w =\frac{1}{2 \pi} \int_{-\infty}^{\infty} d k_{z} \partial_{k_{z}} \phi
=\frac{1}{2} (w_1+w_2)\,,
\end{equation}
where
\begin{equation}
w_1 =\frac{1}{2 \pi} \int_{-\infty}^{\infty} d k_{z} \partial_{k_{z}} \phi_1\,
\end{equation}
\begin{equation}
w_2 =\frac{1}{2 \pi} \int_{-\infty}^{\infty} d k_{z} \partial_{k_{z}} \phi_2,
\end{equation}
with
$$
\tan \phi_{1}=\frac{\operatorname{Re}\left(h_{x}\right)+\operatorname{Im}\left(h_{z}\right)}{\operatorname{Re}\left(h_{z}\right)-\operatorname{Im}\left(h_{x}\right)},$$
$$
\tan \phi_{2}=\frac{\operatorname{Re}\left(h_{x}\right)-\operatorname{Im}\left(h_{z}\right)}{\operatorname{Re}\left(h_{z}\right)+\operatorname{Im}\left(h_{x}\right)}.$$

For our model, when $k_\rho^4-8 k_\rho^2-4>0$, $\phi$ is continuous and thus gives a trivial winding number. However, when $k_\rho^4-8 k_\rho^2-4<0$,
$\phi$ is discontinuous at
\begin{equation}
k_c^{p,n}=\pm \sqrt{2 \sqrt{2 k_\rho^2-1}-k_\rho^2}\,.
\end{equation}
At the discontinuities, the two real angles are given by
\begin{equation*}
s_1^{n,\pm}\equiv\phi_{1}(k_{z} \rightarrow k_c^{n,\pm})=\pm \frac{\pi}{2} \operatorname{sgn}[h_x(k_\rho,k_c^{n,\pm})+\gamma_z]\,,
\end{equation*}
\begin{equation*}
s_2^{n,\pm}\equiv\phi_{2}(k_{z} \rightarrow k_c^{n,\pm})=\pm \frac{\pi}{2} \operatorname{sgn}[h_x(k_\rho,k_c^{n,\pm})-\gamma_z]\,,
\end{equation*}
\begin{equation*}
s_1^{p,\pm}\equiv\phi_{1}(k_{z} \rightarrow k_c^{p,\pm})=\mp \frac{\pi}{2} \operatorname{sgn}[h_x(k_\rho,k_c^{p,\pm})+\gamma_z]\,,
\end{equation*}
\begin{equation*}
s_2^{p,\pm}\equiv\phi_{2}(k_{z} \rightarrow k_c^{p,\pm})=\mp \frac{\pi}{2} \operatorname{sgn}[h_x(k_\rho,k_c^{p,\pm})-\gamma_z]\,.
\end{equation*}
Then the winding numbers are given by,
\begin{equation*}
w_1=\operatorname{sgn}[h_x(k_\rho,k_c^{p,\pm})+\gamma_z]-\operatorname{sgn}[h_x(k_\rho,k_c^{n,\pm})+\gamma_z]\,,
\end{equation*}
\begin{equation*}
w_2=\operatorname{sgn}[h_x(k_\rho,k_c^{p,\pm})-\gamma_z]-\operatorname{sgn}[h_x(k_\rho,k_c^{n,\pm})-\gamma_z]\,.
\end{equation*}

\section{Wannier-Stark function estimation}\label{appd}

\begin{figure}[htbp]
	\centering
	\includegraphics[width=\textwidth]{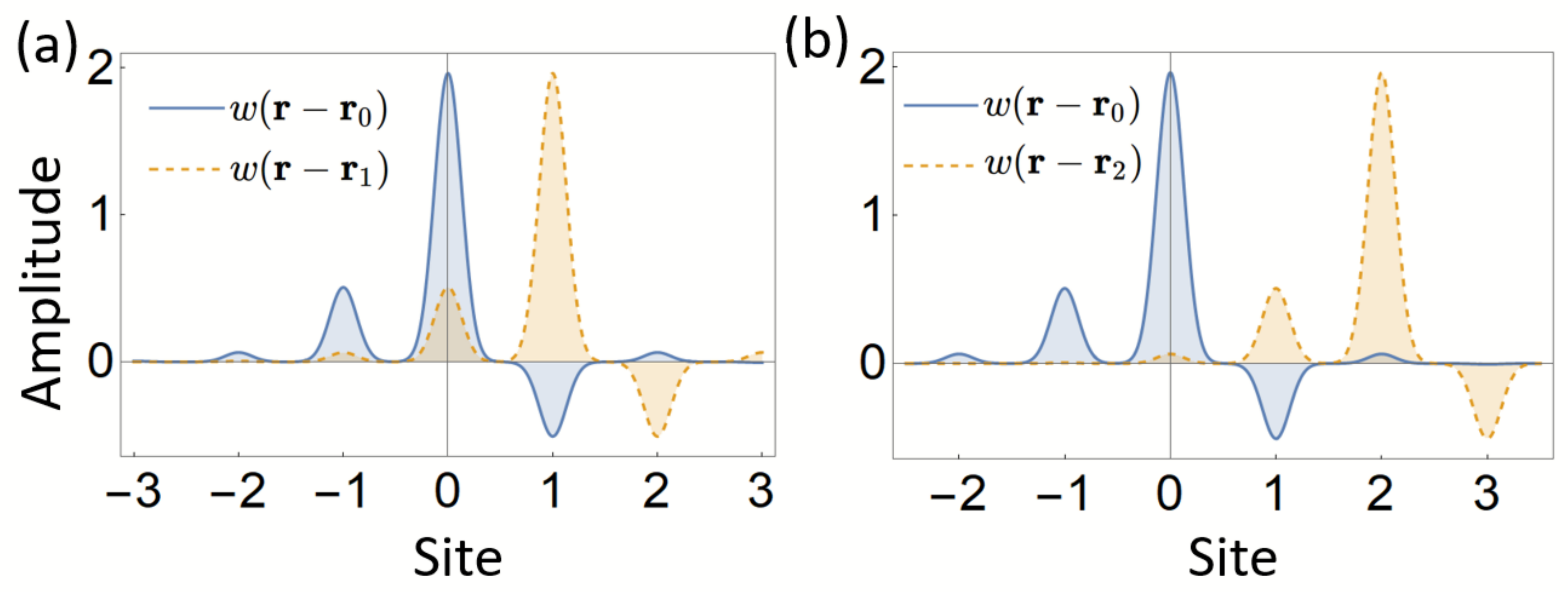}
	\caption{(a) Wannier-Stark function centered at site 0 and site 1. (b) Wannier-Stark function centered at site 0 and site 2. The parameters are chosen as $V_0=2.3 E_r$ and linear tilt $\Delta_x/2\pi=200~\mathrm{Hz}$.}
	\label{fig9}
\end{figure}

For a linear tilted optical lattice, the Wannier-Stark function takes the form \cite{Mitake2013}
\begin{equation}
w_i(\mathbf{r}'-\mathbf{r}_l)=\sum_m J_{m-l}(\frac{2t_0}{\Delta_i})\overline{w}(\mathbf{r}'-\mathbf{r}_l)\,,
\end{equation}
where $i=x,y,z$, $J_{m-l}(x)$ are the Bessel functions of the first kind, and $\overline{w}(\mathbf{r}'-\mathbf{r}_l)$ is the Wannier function centered at the site $\mathbf{r}_l$ and it takes a form of Gaussian function,
\vspace{-0.2 in}
\begin{equation*}
\overline{w}(\mathbf{r})=\frac{1}{\pi^{1/4}\sigma^{3/2}}e^{-\mathbf{r}^2/2\sigma^2}\,,\vspace{0.2 in}
\end{equation*}
with $\sigma=\hbar/[m^{1/2}(4E_RV_0)^{1/4}]$. As discussed in the main text, the Raman-assisted hopping rate is given by
\begin{equation*}
t_{\mathbf{r}, \mathbf{m}}=\frac{\Omega_{\beta \mathbf{m}}^{*} \Omega_{\alpha \mathbf{m}}}{\delta} \int d^{3} \mathbf{r}^{\prime} w^{*}\left(\mathbf{r}^{\prime}-\mathbf{r}-\mathbf{m}\right) e^{i \delta \mathbf{k} \cdot \mathbf{r}^{\prime}} w (\mathbf{r}^{\prime}-\mathbf{r} )\,.
\end{equation*}
We can calculate the overlap integral of nth nearest hopping,
\begin{equation*}
\Lambda(n) \equiv \int dx w^*(x+na)w(x)\int dy w^*(y)w(y) \int dz w(z)w(z).
\end{equation*}

We illustrate the Wannier-Stark functions in Fig. \ref{fig9}. One can see that the overlap between two next-nearest centered Wannier-Stark functions $\Lambda(2)$ has the same order of $\Lambda(1)$ due to tilting the lattice. For nature tunneling rate $t_0/\hbar=2\pi\times 50~\mathrm{Hz}$ and $\Delta_x/2\pi=200~\mathrm{Hz}$, $\Lambda(2)\approx 0.117$ while $\Lambda(1)\approx 0.34$.

\end{appendix}

\end{document}